\documentclass[a4paper]{memoir}
\chapterstyle{veelo}
\usepackage[showerrors,table,dvipsnames*,hyperref]{xcolor}
\usepackage{ucs}
\usepackage[utf8x]{inputenc}
\usepackage{inputenc}
\usepackage{color,graphicx}
\usepackage[english]{babel}
\usepackage{amsmath}
\usepackage{amssymb}
\usepackage{amsxtra}
\usepackage{listings}
\usepackage{verbatim}
\usepackage[square,comma,sort&compress]{natbib}
  \makeatletter
    \let\c@lofdepth\relax
    \let\c@lotdepth\relax
  \makeatother
\usepackage{subfigure}
\usepackage{listings}
\usepackage{slashbox}
\usepackage{ctable}
\usepackage{rccol}
\addto\captionsenglish{%
}
\newcommand{\skiptwo}{\multicolumn{2}{c}{}}

\DeclareFontFamily{U}{euc}{}
\DeclareFontShape{U}{euc}{m}{n}{<-6>eurm5<6-8>eurm7<8->eurm10}{}%
\DeclareSymbolFont{AMSc}{U}{euc}{m}{n} 
\DeclareMathSymbol{\umu}{\mathord}{AMSc}{"16}

\newcommand\cellavision{CellaVision\texttrademark}

\makerunningwidth{companion}{\headwidth}
\title{Classification of Cell Images Using MPEG-7-influenced Descriptors and Support Vector Machines in Cell Morphology}
\date{\today}
\author{Tobias Abenius\\
{\normalsize tobbe@tobbe.nu}
}
\setsecnumdepth{subsection}

\DeclareMathOperator*{\argmax}{arg\ max}
\DeclareMathOperator*{\argmin}{arg\ min}
\newcommand{\vect}[1]{\boldsymbol{\mathbf{#1}}}
\newcommand{\trans}[1]{{#1}^{\ensuremath{\mathsf{T}}}} 
\newcommand{\partder}[2]{\frac{\partial #1}{\partial #2}} 
\providecommand{\abs}[1]{\lvert#1\rvert}
\providecommand{\norm}[1]{\lVert#1\rVert}
\newcommand{\field}[1]{\mathbb{#1}}

\newcommand{\R}{\field{R}}
\newcommand{\Z}{\field{Z}}
\newcommand{\N}{\field{N}}
\newcommand{\x}{\ensuremath{\vect x}}
\newcommand{\w}{\ensuremath{\vect w}}
\newcommand{\X}{\mathcal{X}}
\newcommand{\Y}{\mathcal{Y}}
\newcommand{\sgn}{\ensuremath{\mathrm{sgn}}}

\newenvironment{vardesc}{%
  \settowidth{\parindent}{Where: }
  \makebox[0.92\parindent][r]{Where: }}{\vspace{5pt}\\}

\begin{document}
\input epsf.tex
\bibliographystyle{plain}
\numberwithin{equation}{chapter}
\numberwithin{figure}{chapter}
\numberwithin{table}{chapter}

\newcommand{\puttitle}{\begin{center}\Large \thetitle\end{center}}
\newcommand{\putTitle}{\begin{center}\huge \thetitle\end{center}}

\newcommand\csthesistext{\begin{minipage}[r]{1.5\textwidth}
\large%
Examensarbete för 30 hp\\
Institutionen för datavetenskap, Naturvetenskapliga fakulteten, Lunds universitet\\
\ \\
Thesis for a diploma in Computer Science, 30 ECTS credits\\
Department of Computer Science, Faculty of Science, Lund University
\end{minipage}
}
  \copypagestyle{firstpage}{empty}
	\makeoddfoot{firstpage}{\csthesistext}{}{}
  \makerunningwidth{firstpage}{0.85\headwidth}
\thispagestyle{firstpage}
	\calccentering{\unitlength}
	\begin{adjustwidth*}{\unitlength}{-\unitlength}
		\putTitle
		\vspace{1em}
		\begin{center}
			\large \theauthor
		\end{center}
		\flushbottom
		\vfill
		\hspace{-1.5cm}%
\begin{minipage}[r]{1.5\textwidth}
	\raggedright 
\end{minipage}

	\end{adjustwidth*}
\clearpage

\thispagestyle{empty}
	\calccentering{\unitlength}
	\begin{adjustwidth*}{\unitlength}{-\unitlength}
		\puttitle
\vspace{1em}
\abstractrunin
\abslabeldelim{---}
\setlength{\absparindent}{0pt}
\setlength{\abstitleskip}{-\absparindent}
\begin{abstract}%
\footnotesize%
Counting and classifying blood cells is an important diagnostic tool in medicine. 
Support Vector Machines are increasingly popular and efficient and could replace artificial neural network systems.
Here a method to classify blood cells is proposed using SVM.
A set of statistics on images are implemented in C++.
The MPEG-7 descriptors {\em Scalable Color Descriptor}, 
{\em Color Structure Descriptor}, {\em Color Layout Descriptor} and {\em Homogeneous Texture Descriptor} are extended in size and combined with textural features corresponding to textural properties perceived visually by humans. 
From a set of images of human blood cells these statistics are collected.
A SVM is implemented and trained to classify the cell images. 
The cell images come from a \cellavision~DM-96 machine which classify cells from images from micro\-scopy.
The output images and classification of the \cellavision\ machine is taken as ground truth, a truth that is 90-95\% correct.
The problem is divided in two --- the {\em primary} and the {\em simplified}. 
The primary problem is to classify the same classes as the \cellavision\ machine. 
The simplified problem is to differ between the five most common types of white blood cells.
An encouraging result is achieved in both cases --- error rates of 10.8\% and 3.1\% --- considering that the SVM is misled by the errors in ground truth.
Conclusion is that further investigation of performance is worthwhile.
\end{abstract}
\vfill
\begin{center}
	\Large Klassificering av cellbilder med hjälp av MPEG-7-inspirerade mått och support vector machines i cellmorfologi
\end{center}
\vspace{1em}
\renewcommand{\abstractname}{Sammanfattning}
\begin{abstract}%
\footnotesize%
Att räkna och klassificera blodceller är ett viktigt dia\-gnost\-iskt redskap inom läkarvetenskapen.
Support Vector Machines är ef\-fekt\-iva, ökar i popularitet och kan ersätta artificiella neurala nätverkssystem.
Här föreslås en metod för att klassificera blodceller m.h.a. SVM.
En mängd stat\-istika på bilder implementeras i C++.
De s.k. MPEG-7 descriptors {\em Scalable Color Descriptor}, 
{\em Color Structure Descriptor}, {\em Color Layout Descriptor} och {\em Homo\-geneous Texture Descriptor} utvidgas i storlek och kombineras med textur-mått motsvarande textur-egenskaper som uppfattas visuellt av människor.
Från en mängd bilder av mänskliga blodceller samlas dessa mått.
En SVM im\-ple\-ment\-eras och tränas att klassificera cellbilderna.
Cellbilderna kommer från en \cellavision~DM-96 som klassificerar celler från mikroskoperade bilder.
Bilderna och dess klasser från en \cellavision~DM-96-maskin tas som facit, ett facit som är 90-95\% korrekt.
Problemet delas i två --- det {\em primära} och det {\em förenklade}.
Det primära problemet är att skilja mellan de klasser som \cellavision s maskin gör.
Det förenklade problemet är att skilja mellan de fem vanligaste typerna av vita blodkroppar.
Ett glädjande resultat uppnås i båda fallen --- felfrekvenser om 10,8\% och 3,1\% --- med tanke på att SVM missleddes av felen i det tagna facitet.
Slutsatsen är att vidare studier angående prestanda är lönsamma.
\end{abstract}

	\end{adjustwidth*}

\clearpage
\thispagestyle{empty}

\hspace{10cm}\begin{minipage}[r]{100pt}\vspace{3cm}
	\raggedleft \huge \textit{to Britta,}\protect\\\textit{to my family}
\end{minipage}

\cleardoublepage

\settocdepth{subsection}
\tableofcontents{}
\listoftables{}

\chapter*{Acknowledgments\markboth{Acknowledgments}{}}
\addcontentsline{toc}{chapter}{Acknowledgments}
First of all thanks to Doc. Christian Balkenius and Doc. Jacek Malec for supervising my thesis. 
To Dr. Ferenc Belik for managing all practical details.
to Doc. B.S. Manjunath for inspiration and lending me figure~\ref{fig:freqtouch}. 
To Sebastian Ganslandt for initial chats about support vector machines and thesis ideas. 
To all others that made this work possible.

\chapter{Introduction}
\pagenumbering{arabic}
After the introduction of MPEG-7 descriptors by the Movie Producers Expert Group (MPEG) committee\cite{manjunath2001ColorTextureDescriptors} it is interesting to see how these features perform in the field of machine learning. In this thesis a subset of them will be tested on the problem of classifying different cell types, i.e. {\em cell morphology}, by using Support Vector Machines.

In medicine, more specifically the fields of hematology and infectious diseases, classifying different kinds of blood cells can be used as a tool in diagnosis --- by counting certain cells' relative frequencies and comparing to what is normal, conclusions can be made about possible diagnosis.
\begin{table}
\begin{tabular}{rl}
\toprule
{\scshape Type} & {\scshape Approx. Abundance}\\
\midrule
neutrophil granulocytes & 70\%  \\
eosinophil granulocytes & 1-6\% \\
basophil granulocytes & 0.01-0.3\%  \\
lymphocyte & 20-40\% \\
monocytes & 3-8\% \\
\bottomrule
\end{tabular}
\caption{Abundance of different types of white blood cells (leukocytes) in healthy humans}
\label{tab:cellabundance}
\end{table}
\begin{figure}
	\begin{center}
		\subbottom[Neutrophil Granulocyte, segmented (class 1)]{\includegraphics[scale=0.6]{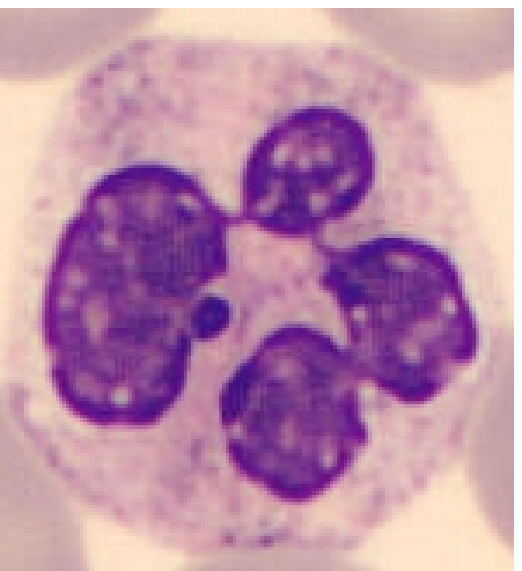}}
		\hfill
		\subbottom[Neutrophil Granulocyte, band (class 6)]{\includegraphics[scale=0.6]{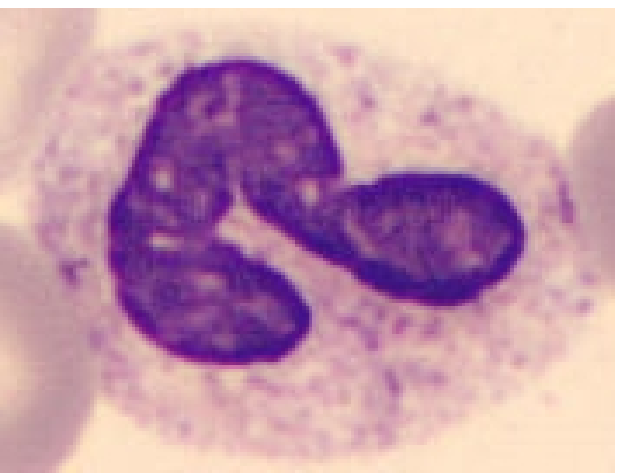}}
		\hfill
		\subbottom[Eosinophil Granulocyte (class 2)]{\includegraphics[scale=0.6]{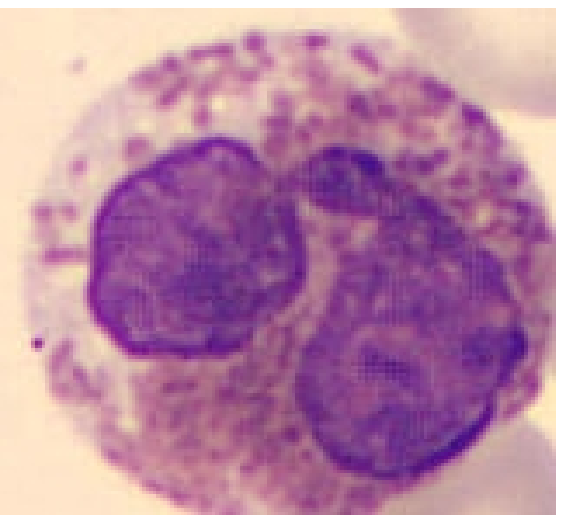}}
		\hfill
		\subbottom[Basophil Granulocyte (class 3)]{\includegraphics[scale=0.6]{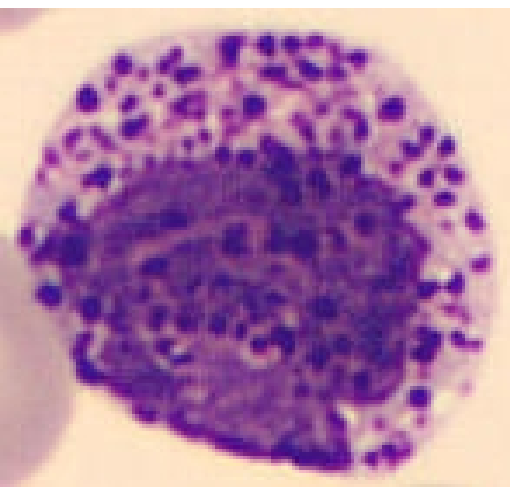}}
		\hfill
		\subbottom[Lymphocyte (class 4)]{\includegraphics[scale=0.6]{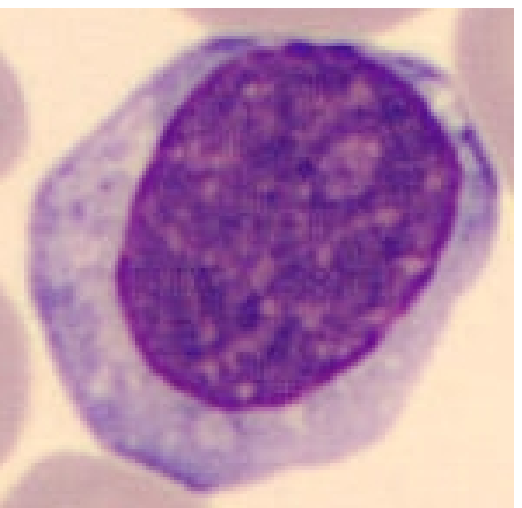}}
		\hfill
		\subbottom[Monocyte (class 5)]{\includegraphics[scale=0.6]{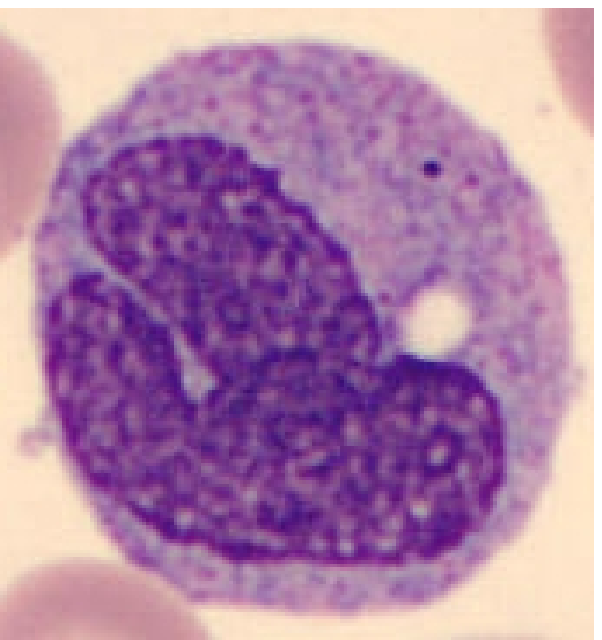}}
		\caption{Some typical images of common white blood cells}
		\label{fig:celltypes}
	\end{center}
\end{figure}

Classifying cells using microscopy is used to classify infectious diseases by determining the relative amount of cells called neutrophils compared to the amount of cells called lymphocytes. 
Typical relative frequencies of the cells are found in table~\ref{tab:cellabundance}. 
Typical images of some common cells are found in figure~\ref{fig:celltypes}.

Another method used is flow cytometry where receptors on the cells are colored and the different types of cells are counted. Flow cytometry uses a complicated and expensive apparatus while microscopy is very cheap. 

However, microscopy is personnel intensive,
many cells are hard to classify even for human experts, 
often several experts are needed to be certain.
To be able to classify cells, great efforts of training are required, even more, to sustain competence, regular frequent work is required.
This competence is impossible to sustain at small clinics or in the countryside especially in developing countries. Instead, samples have to be sent to hematology labs.

As processing power becomes cheaper and machine learning and computer vision algorithms grow better, machines can help less experienced personnel or give preliminary results while waiting for definite results.

The problem this thesis try to investigate is how well these different types of white blood cells can be classified using a Support Vector Machine and a set of measures on the images, called {\em features}.

There has been a lot of hype about Support Vector Machines since its introduction in the 1990's. 
SVM is applied within a broad range of fields, 
from
bioinformatics\cite{LengauerBioinformaticsPredictionOfHIVCoreceptorUsage} to
food engineering\cite{DuMultiClassificationOfPizzaUsingComputerVisionAndSVM},
iris recognition\cite{IrisRecognitionBasedOnScoreLevelFusionByUsingSVM}, 
texture classification and object recognition\cite{ZhangLocalFeaturesAndKernelsForClassificationOfTextureAndObjectCategories}.
It is now one of standard tools available for machine learning%
---A recent search for ``Support Vector Machine'' (SVM) gave 6\,394 articles compared to 17\,893 for ``Artificial Neural Network'' (ANN) which has existed for much longer. That is why my supervisor and I chose to work with SVM.

The SVM is trained with measures of the cell images, called {\em features} or {\em descriptors}. These are values that describe the essence of an image.
In this thesis I will describe and implement a subset of the color and texture descriptors found in the MPEG-7 standard with minor variance.
I chose to work with MPEG-7 as a guide because of the MPEG committee's well known expertise. 

The MPEG committee developed e.g. the audio compression techniques used in MPEG-1 Layer 3 (MP3), the video compression used in e.g. DVDs (MPEG-4) and MPEG-7. The committee consists of experts from a broad range of areas that deal with digital information.\cite{urlMPEG7}

MPEG-7 identify several descriptors which has proved useful in the {\em Color and Texture Core Experiments}\cite{manjunath2001ColorTextureDescriptors} while developing of the standard. 
They have proved useful in image browsing, search and retrieval%
\cite{manjunath2000ATextureDescriptorForBrowsingAndSimilarityRetrieval}%
\ as well as in image classification%
\cite{SpyrouFuzzySVMForImageClassificationFusingMPEG7VisualDescriptors}%
.
Color histogram based features has been successful both in image retrieval%
\cite{SergyanColorHistogramFeaturesBasedImageClassificationInContentBasedImageRetrievalSystems} and image classification%
\cite{SergyanColorHistogramFeaturesBasedImageClassificationInContentBasedImageRetrievalSystems,ChapelleSVMForHistogramBasedImageClassification,BarlaOldFashionedStateOfTheArtImageClassification}%
\ systems.
Texture features like {\em Gabor Wavelet Filter Bank} used in MPEG-7 has been successfully applied to\ %
iris%
\cite{IrisRecognitionBasedOnScoreLevelFusionByUsingSVM} and 
facial expression%
\cite{BuciuICAAndGaborRepresentationForFacialExpressionRecognition}%
\ recognition.

%


\chapter{Background}
\section{Support Vector Machines}
In this section I will briefly introduce Support Vector Machines from a theoretical perspective. 
%
Further introduction may be found in Bishop's book\cite[chapters 6,7 and E]{Bishop}.
If more substance is wanted I recommend reading the whole book by Christianini and Shawe-Taylor\cite{Nello}. 
The very thorough coverage of the topic by its original implementor Vapnik in his book\cite{Vapnik}, sometimes called the bible, was often an additional useful source for me.

\subsection{Supervised Learning}
{\em Supervised learning} is a kind of machine learning where the machine is fed with {\em examples}, i.e. instances of data tied to their class.
The machine is told what class an instance belongs to.

The task that a learning machine performs is to recognize an element $\x \in \X$ as a member of a class --- to classify it.
These classes are called destination values and I use the notation $y\in\Y$. In the binary case for example $\Y=\{-1,+1\}$.
The task would then be to construct a function such that $d(\vect x,\vect \alpha)=y$, given $\vect \alpha$ is the information the machine has previously gathered during the {\em training} process.
During training, the machine observes a tuple of pairs
\begin{align*}
S = \big((\vect{x}_1,y_1),\ldots,(\vect{x}_\ell,y_\ell)\big) \subseteq (\X \times \Y)^\ell,
\end{align*}
which is called the {\em training set}, and produces parameters $\vect \alpha \in \R^n$ deduced from this information.%
\cite{Nello}

\subsection{Linear Learning Machines}
Imagine the space $\X$ which has $n$ dimensions. To be able to classify instances into the two classes labeled {\em positive}, $y=+1$, or {\em negative}, $y=-1$, a hyperplane, i.e. an affine subspace of dimension $n-1$, must be found that separates the instances of the respective classes from each other. If such a hyperplane exists, the data is said to be {\em linearly separable}.

Imagine a two-dimensional coordinate system in which the instances are placed. If a straight line can be placed between the two classes of instances, the data is linearly separable. That straight line is a hyperplane of dimension 1. The generalized hyperplane of dimension $n-1$ is defined by the equation
\begin{align*}
	\langle \w,\x \rangle + b = 0.
\end{align*}
The normal vector \w\ is orthogonal to the hyperplane and the bias $b$ is the hyperplane's offset from the origin.

Now consider the function
\begin{align}
	f(\x) = \langle \w,\x \rangle + b = \sum_{i=1}^n w_i x_i + b
\end{align}
\nopagebreak%
\begin{vardesc}%
\x\ -- instance

\w\ -- coefficients learned

$b$ -- system bias
\end{vardesc}
It will tell whether an instance is above or below the hyperplane.
This is similar to {\em linear regression} in statistics.

A decision function for the binary classification case then becomes
\begin{align*}
	d(\x) &= \sgn(f(\x))\\
	\sgn(a) &= \begin{cases}
		-1,&\;a < 0\\
		+1,&\;a \ge 0
	\end{cases}
\end{align*}

An example of an iterative algorithm that find the vector $\w$ from a set of $\x \in \X$ is Rosenblatt's {\em perceptron} which was the first and simplest type of an Artificial Neural Networks (ANN). It is guaranteed to converge if the data is linearly separable. 
This criterion could also be written
\begin{align*}
	\label{eq:linear_separable}
	\exists \w \forall i: \gamma_i &= y_i(\langle \w,\x_i\rangle + b) > 0,\\
i \in [0,\ell),
\end{align*}
i.e. all instances are classified correctly. The quantity $\gamma_i$ is called the {\em margin} as it specifies how far from the hyperplane an instance is. 
If \w\ and $b$ are normalized, to $\frac{\w}{\norm{\w}}$ and $\frac{b}{\norm{\w}}$, then the margin is called the {\em geometric margin} which measures the euclidean distances of the points \x\ to the hyperplane. 
The closest point, the $\x_i$ with minimal $\gamma_i$, define the {\em margin of a hyperplane} which is a stripe of empty space where no instances are. If the data is not linearly separable $\exists i: \gamma_i \le 0$.%
\cite{Nello,Bishop}

\subsection{Maximum Margin Classifier}
The task of a {\em maximum margin classifier} is to maximize the margin which can be motivated, using {\em statistical learning theory}, gives the least generalization error.

The maximum margin solution, the optimal \w\ and $b$, is found by solving
\begin{align*}
	\argmax_{\w,b} \left\{\min_i \frac{\gamma_i}{\norm{\w}}\right\} = \argmax_{\w,b} \left\{\frac{1}{\norm{\w}}\min_i y_i(\langle \w,\x_i\rangle + b)\right\}
\end{align*}
To solve this first rescale $\w \rightarrow \kappa\w$ and $b \rightarrow \kappa b$. The distance to the hyperplane is still the same $\min_i \gamma_i$. Then set
\begin{align*}
	\gamma_j = y_j(\langle \w,\x_j\rangle+b = 1
\end{align*}
for the point $\x_j$ that is closest to the hyperplane. 
All points will then have $\gamma_i \ge 1$ and since the minimum $\gamma_j = 1$ all that have to be done is to maximize $\norm{\w}^{-1}$ or minimize $\norm{\w}^2$. The problem that is left is to
\begin{equation}%
	\label{eq:margin}\begin{split}
		\text{find} &\quad \argmin_{\w,b} \frac{\norm{\w}^2}{2},\\*
		\text{subject to} &\quad \gamma_i \ge 1,
	\end{split}%
\end{equation}
which is much easier. This problem is what is called a {\em quadratic programming problem} and can be solved using the theory of optimization theory and {\em Lagrange Multipliers}.%
\cite{Nello,Bishop}

\subsection{Optimization Theory}
The theory on Lagrangian multipliers states that to
\begin{equation*}\begin{split}
	\text{optimize} \quad & f(\x) \\
	\text{subject to} \quad & g(\x) \ge 0\\
\end{split}\end{equation*}
	{one should optimize the Lagrangian function}
\begin{equation*}\begin{split}
	&L(\x,\alpha) = f(\x)+\alpha g(\x)\\
	\text{subject to} \quad &	g(\x) \ge 0\\
	&\alpha \ge 0\\
	&\alpha g(\x) = 0.
\end{split}\end{equation*}
These conditions are known as the {\em Karush-Kuhn-Tucker}(KKT) conditions.
More generally, to add more constraints $g_j(\x)$, replace the $\alpha g(\x)$ with a linear combination of all Lagrange multipliers $\alpha_j$ and their corresponding functions $g_j(\x)$\cite{Bishop}:
\begin{equation*}\begin{split}
	\text{optimize}\quad\qquad\! & L(\x,\{\alpha_j\}) = f(\x) + \sum_{j=1}^{J}\alpha_j g_j(\x)\\
\text{subject to} \quad \forall j:\ &	g_j(\x) \ge 0\\
 &\alpha_j \ge 0\\
 &\alpha_j g_j(\x) = 0.
\end{split}\end{equation*}

In order to quickly find a solution to \eqref{eq:margin} it can now be rewritten as the Lagrangian function
\begin{align*}
	\label{Lagrangian}
	L(\w,b,\alpha) = \underbrace{\frac{1}{2}\norm{\w}^2}_{f(\x)} - \sum_{i=1}^\ell \alpha_i \underbrace{(y_i(\langle \w,\x_i\rangle + b) - 1)}_{g_i(\x)}.
\end{align*}
The constraint function is negative because we are minimizing wrt $\norm{\w}$ and $b$ while maximizing wrt $\vect \alpha$. 
To finally arrive at what is called the {\em dual representation} of the maximum margin problem the derivatives of $L$ wrt to \w\ and $b$, are set to $0$.
Maximizing this dual representation,
\begin{equation}\begin{split}
	& W(\vect \alpha) = {\tilde{L}}(\vect\alpha) = \sum_{i=1}^{\ell} \alpha_i - \frac{1}{2}\sum_{i=1}^{\ell} \sum_{j=1}^{\ell} \alpha_i \alpha_j y_i y_j \langle \x_i, \x_j \rangle, \\
	\text{by finding} \quad & \vect \alpha,\\
	\text{subject to} \quad &  \forall i: \alpha_i \ge 0, \\
	&            \sum_{i=1}^{\ell} \alpha_i y_i = 0,
\end{split}\end{equation}
will construct the maximal margin classifier.%
\cite{Nello,Bishop,Vapnik}

The instances that have a corresponding $\alpha_i > 0$ are called {\em support vectors}. That is because they lie on the margin. They are thus used in the decision function.

Note how the input variables $\x_i$ are only used in an inner product which let the SVM avoid the {\em curse of dimensionality} caused by a data set with instances of too high dimension.%
\cite{Nello}

\subsection{The Kernel Trick}
The Kernel Trick is used implicitly in Support Vector Machines but it has also been tried out in e.g. RBF Networks, which is a kind of ANN.\cite{Bishop}

The inner product used in the dual optimization problem can be a linear one. 
Though it will not separate the instances fully when the dataset is not linearly separable, data must be mapped to another space where it is.

A non-linear {\em feature function} $\phi(\x)$ can do such a mapping. 
However, there is no need to know the feature function explicitly, it is easier to define it implicitly via a {\em Mercer Kernel}.\cite{Nello}

A complete, normed space with an inner product is called a {\em Hilbert Space}
One of the beauties of Hilbert spaces lies in that any given function in the $L_2$ space could be approximated infinitely well in the $\norm{\cdot}_2$ and represented by an infinite linear combination of some coefficients and some basis functions. 
An example of this is the {\em Fourier Series} using Fourier coefficients and the Dirichlet Kernel Functions $\{e^{-ikx}\}_k$.

A special kind of Hilbert spaces are the ones which are called {\em Reproducing Kernel Hilbert spaces}. A function $\langle \x_i,\x_j\rangle = K(\x_i,\x_j) = \phi(\x_i)\phi(\x_j)$ is called a kernel when it satisfies the criteria in Mercer's Theorem.

A Mercer kernel $K$ is defined as an inner product on elements of some space $\mathcal X$.\cite{Nello}
An inner product is a function that is a positive-definite sesqui-linear\footnote{anti-linear in the second argument and linear in the first} form. 
In the $\R$ case this becomes
a function
\begin{align}
\notag \langle\cdot,\cdot\rangle &: \mathcal X \times \mathcal X \rightarrow \R\\
\intertext{such that}
K(\vect{x},\vect{z}) &=\langle \vect{x} , \vect{z} \rangle = {\langle \vect{z} , \vect{x} \rangle} = {K(\vect{z},\vect{x})} & \tag{Symmetry}\\
K(a\vect{x}+b\vect{y},c\vect{z}) &= ab{c}\big(K(\vect{x},\vect{z})+K(\vect{y},\vect{z})\big) & \tag{Bilinearity}\\
\forall \vect{x}: K(\vect{x},\vect{x}) &\ge 0 & \tag{Positivity}\\
K(\vect{x},\vect{x}) &= 0 \iff \vect{x} = \vect{0}& \tag{Definiteness}\\
\intertext{A Mercer kernel also have non-negative eigenvalues $\lambda_i$ of the Gram matrix $\vect G$ since it's defined as a Hermitian matrix}
\forall i:  \lambda_i &\ge 0 | \vect{G}&\tag{Positive semi-definite Gram matrix}\\
\vect{G} &= \Big(K\big({\vect x}_i,{\vect x}_j\big)\Big)_{i,j \in [1,\ell]^2}\label{eq:GramMatrix}
\end{align}
Note that the elements of the space $\mathcal X$ do not need to be real vectors as they will be in this context, they could also be e.g. strings of symbols as well. 
As soon as a symmetric sesqui-linear positive-definite function could be defined on the elements of the space $\mathcal X$, the space becomes an {\em inner product space} and the Support Vector Machine will do its job.\cite{Nello}

Here are some commonly used Mercer kernels defined on $\R^n\times\R^n$%
\cite{Nello,Bishop,Vapnik}:
\begin{align}
	\label{eq:kernelexamples}
\langle \vect x, \vect y\rangle_{Linear} &= \trans{\vect{x}}\vect{y} & \tag{Linear, dot product, kernel}\\
\langle \vect x,\vect y\rangle_{Poly}    &= \Big(\trans{\vect{x}}\vect{y} + 1\Big)^d & \tag{Complete Polynomial of degree $d$}\\
\langle \vect{x},\vect{y}\rangle_{RBF}   &= \exp\left(-\frac{1}{2\sigma^2}\norm{\vect{x}-\vect{y}}\right)  & \tag{Gaussian, Radial Basis Function}\\
\langle \vect x, \vect y\rangle_{MLP}    &= \tanh(\trans{\vect{x}}\vect{y}+b) & \tag{Multilayer perceptron, for some $b$}
\intertext{the norm used in RBF is usually the euclidean distance, $p=2$ below}
\norm{x - y}_{L^p} &= \Big(\sum_{i}\abs{x_i - y_i}^p\Big)^{1/p} &\tag{$L^p$ distance}
\end{align}

\subsection{Gradient Ascent}
An easy approach to find coefficients $\vect \alpha$ is to update them in the direction of the gradient of the objective function $W(\vect \alpha)$,
\begin{align*}
	\partder{W(\vect \alpha)}{\alpha_i} &= 1 - y_i \sum_{j=1}^{\ell} \alpha_j y_j \langle\x_i,\x_j\rangle.
	\intertext{To maximize the objective function $W(\vect \alpha)$ one could just iterate}
	\alpha_i' &\leftarrow \alpha_i + \eta \partder{W(\vect \alpha)}{\alpha_i}.
\end{align*}
\nopagebreak%
\begin{vardesc}%
	$\eta$ -- the learning rate
\end{vardesc}
It is shown e.g. in Nello's book that setting $\eta=\frac{1}{K(\x_i,\x_j)}$ maximizes the gain if the $\alpha_i \in [0,C], C\in \R$ and that convergence is guaranteed if the hyperplane exists.\cite{Nello}

\subsection{Multiclass SVM}
There are three major methods for training a set of classifiers to be able to classify several classes\cite{HsuLinMultiClass}, i.e. $|\Y| = k > 2$.

In the {\em one-against-the-rest} method $k$ binary classifiers are created where classifier $i \in [0,k)$ is told that all examples with class $i$ are positive and the rest are negative. 
When predicting which class $\vect x$ belongs to all classifiers are tested and the one which gave the highest certainty wins.

In the {\em one-against-one} method $k(k-1)/2$ binary classifiers are created such that all 2-combinations of classes $i,j$ have a corresponding classifier.
\begin{align*}
	C^n_2 = {n \choose 2} = \frac{n!}{2!(n-2)!} = \frac{n(n-1)(n-2)!}{2(n-2)!} = \frac{n(n-1)}{2}
\end{align*}
The prediction is then done by voting, all binary classifiers vote on their respective class $i$ or $j$. The class with the highest vote wins, this approach is called the "Max Wins" strategy.

{\em Direct Acyclic Graph SVM} (DAGSVM) is the third method. It uses the same training method as one-against-one but a different decision mechanism. The classifiers are placed in a rooted DAG with the classifiers as internal nodes and the classes as leaves. Starting at the root a binary decision means move either left or right. When a leaf is reached the decision is done.%
\cite{HsuLinMultiClass}

\section{Features}
Features, or descriptors, try to take useful information out of an image --- color distribution, measures on edges and texture properties. 
They capture information in a more condensed and efficient way than by just using the color values in each pixel. 

These descriptors are also {\em scale invariant} --- it does not matter which size the images have. 
This is necessary as the images have different sizes.

{\em Scalable Color Descriptor}, {\em Color Structure Descriptor} and {\em Color Layout Descriptor} are the three color descriptors that I describe below and that are implemented in the project.
After the description of those come descriptions of two texture descriptors.
One of them is similar to the {\em Homogeneous Texture Descriptor} from MPEG-7. Another set of descriptors, named {\em Visual Texture Features}, is from an article by Amadasum and King which describe computational measures which approximate how humans perceive texture.\cite{Amadasun1989TexturalFeaturesCorrespondingToTexturalProperties}

\subsection{Scalable Color Descriptor}
\label{sec:scalablecolor}
The HSV space is uniformly quantized into a 3D histogram of 256 bins. Hue is divided into 16 levels, Saturation into 4 and Value into 4. 
In the MPEG-7 specification the $16\times4\times4=256$ bins are truncated to a 11-bit integer mapped to a non-linear 4-bit representation and then encoded using a Haar transform to drastically reduce space footprint. The scalability in this descriptor comes from the ability to choose how many Haar coefficients to store, see an article by Manjunath et al. for more details.\cite{manjunath2001ColorTextureDescriptors}
\subsection{Color Structure Descriptor}
\label{sec:ColorStructureDescriptor}
\begin{figure}
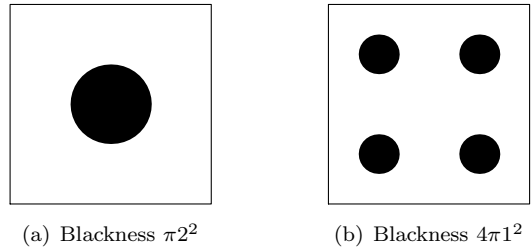

	\begin{center}
	\subbottom[Blackness $\pi 2^2$]{
		\label{fig:colorstructurea}
		\epsfbox{blackness.1}
	}
	\hspace{1cm}
	\subbottom[Blackness $4\pi 1^2$]{
		\label{fig:colorstructureb}
		\epsfbox{blackness.2}
	}
	\caption{These images contain the same amount of black and would yield an identical color histogram but a different color structure descriptor.}
	\label{fig:colorstructure}
\end{center}
\end{figure}

To express local color structure in an image this descriptor slides an $8\times8$-structuring element across the image counting in how many of these elements each color exists. 
By this technique one can differ between the images in figure~\ref{fig:colorstructure}. 

This descriptor is scale invariant as the structuring elements spatial extent scale with the image size. 
The structure element uses replacement sub-sampling if the image is larger than $256\times256$ pixels. 
If e.g. a $512\times512$ image is processed every other row and column will represent the image and the rest of the $2\times2$ areas are thrown away. 
More generally
\begin{align}
p &= \max\{0, \mathrm{round}(0.5\log_2(WH) - 8)\}\\
K &= 2^p,\,\, E=8K
\end{align}
\nopagebreak%
\begin{vardesc}$E\times E$ -- spatial extent of the structuring element

$K$ -- sub-sampling factor
\end{vardesc}
Each bin in the generated histogram represents the number of occasions a structuring element is found to contain the color associated with the bin. 

\subsection{Color Layout Descriptor}
This is kind of a low-pass filter capturing spatial information.
Again it is inspired by the MPEG-7 specification.
The image is first divided in $8\times 8$ blocks. Then interpolation sub-sampling\footnote{the average of all pixels involved in the block represent the whole block as opposed to replacement sub-sampling where a single pixel represent the whole block} is applied, i.e. calculating the average color in each block, giving one representative color for each block.
A 2D discrete cosine transform (DCT-II) is performed on the resulting $8\times 8$ matrix. Low-frequency coefficients are selected using zigzag scanning order, see figure~\ref{fig:zigzag}.
In MPEG-7 the 6 first Y, the 3 first of U and V coefficients are extracted.
\begin{figure}
\begin{center}
\begin{align*}
\begin{pmatrix}
1 & 3& 4 & 10 & 11 \\
2 & 5& 9 & 12 & 19 \\
6 & 8& 13& 18 & 20 \\
7 & 14&17& 21 & 24 \\
15& 16&22& 23 & 25 \\ 
\end{pmatrix}
\end{align*}
\end{center}
\caption{Zigzag scan order of a $5\times 5$ matrix}
\label{fig:zigzag}
\end{figure}

\subsection{Homogeneous Texture Descriptor}
{\em Gabor wavelets} have proved to be the best set of features compared to {\em pyramid-structured wavelet transform} (PWT), {\em tree-structured wavelet transform} (TWT) and {\em multi-resolution simultaneous autoregressive model} (MR-SAR) based de\-script\-ors.%
\cite{manjunath1996TextureFeaturesForBrowsingAndRetrievalOfImageData} 
They are used in the MPEG-7 Homogeneous Texture Descriptor (HTD).

\begin{figure}
	\begin{center}
\epsfbox{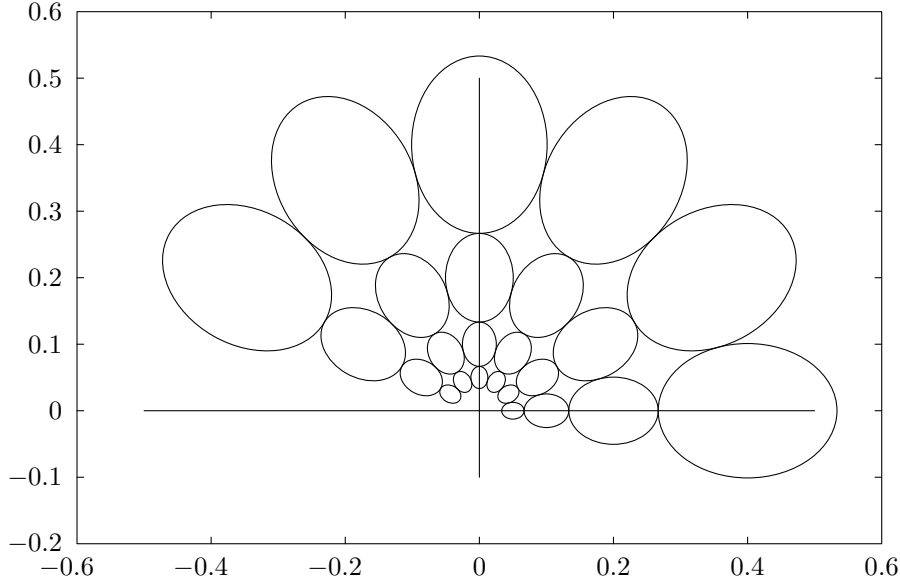}
\caption{$U_{hi} = 0.4, U_{lo} = 0.05, S=5, K=6$}
\label{fig:freqtouch}
\end{center}
\end{figure}

Gabor wavelets are a family of modulated Gaussians, they form a complete basis set implying that, 
any given function $f(\cdot,\cdot)$ can be expanded in terms of these basis functions.
However, as they are not orthonormal, there is redundant information present in a set of coefficients.
To decrease that redundancy I follow the strategy used by Manjunath et al., 
that is aligning the Gaussians such that their half-peaks meet like in figure~\ref{fig:freqtouch}.\cite{manjunath2000ATextureDescriptorForBrowsingAndSimilarityRetrieval} 
.
To achieve this we first make a change of variables. The Gaussian is a Gaussian in both frequency and space domains. The width of the Gaussian in the frequency domain ($\sigma_u, \sigma_v$) is inversely related to the Gaussian in the space domain ($\sigma_x, \sigma_y$). In other words, the wider the Gaussian, the narrower its bandwidth.\cite{wikiGaussianFunction,Wallis_Linear_Models_Of_Simple_Cells_Mammal_Vision_Model}
\begin{align*}
\sigma_x &= \frac{1}{2\pi\sigma_u},\quad\! \sigma_y = \frac{1}{2\pi\sigma_v}\\
\end{align*}
These parameters are needed for scaling
\begin{align*}
a &= (U_{hi}/U_{lo})^{1/(S-1)}, \quad\! \sigma_u = \frac{(a-1)U_{hi}}{(a+1)\sqrt{2\ln 2}},\\
\sigma_v &= \tan\left(\frac{\pi}{2K}\right)\!\left[U_{hi} - 2 \ln 2\left(\frac{\sigma^2_u}{U_{hi}}\right)\middle]\!\middle[2\ln 2 - \left(\frac{(2\ln 2)\sigma_u}{U_{hi}}\right)^{\!\!2\,}\right]^{\frac{1}{2}}\\
\end{align*}
\nopagebreak%
\begin{vardesc}%
$U_{lo}\in \R$ -- lower center frequency of interest

$U_{hi}\in \R$ -- upper center frequency of interest

$m \in [0,S) \subset \Z^+$ -- scale index

$S \subset \N$ -- number of scales

$a > 1 \in \R$ -- scale factor
\end{vardesc}
For different orientations the image needs to be rotated before filtering and scaling wrt $a$.
\begin{align*}
x' &= a^{-m}(x \cos \theta + y \sin \theta)\\
y' &= a^{-m}(-x \sin \theta + y \cos \theta)\\
\theta &= n\pi/K
\end{align*}
\nopagebreak%
\begin{vardesc}%
$n\in [0,K) \subset \Z^+$ -- orientation index

$K\in \N$ -- number of orientations

$\theta \in [0,\pi)$ -- orientation angle
\end{vardesc}

The generated filter bank are matrices that should be convoluted with the image
\begin{align*}
I' = I * G
\end{align*}
\nopagebreak%
\begin{vardesc}%
$*$ -- the convolution operator
\end{vardesc}
See section~\ref{sec:convolution} for details about 2D convolution. 
In figure~\ref{fig:gaborkernels} images of the Gabor wavelet filter bank kernels of different orientations are presented.

In MPEG-7, rotation invariance is achieved in this descriptor, by rotating the features in the direction of the dominant direction. 

\subsection{Visual Texture Features}
The features described in the article by Amadasun and King are implemented. These are features corresponding to properties of texture that humans can perceive. In the article measures of coarseness, contrast, busyness, complexity and strength are introduced and compared by rank with how humans sensed ten natural textures from the widely used Brodatz's album.
I give here a very brief overview of the proposed measures.
They all use a column vector called neighborhood gray-tone difference matrix (NGTDM).\cite{Amadasun1989TexturalFeaturesCorrespondingToTexturalProperties}

\subsubsection{Neighborhood Gray-Tone Difference Matrix}
In a pixel $p$ with coordinates $\langle k,l\rangle$ neighborhood of size $d$, i.e. of the square surrounding a pixel, but without the center pixel the mean is calculated.
\begin{align}
	\label{eq:NGTDMAbar}
	\begin{split}
		\bar{A}_p = \bar{A}(k,l) = \frac{1}{W-1}\left[\sum_{m=-d}^d\sum_{n=-d}^d f(k+m,l+n)\right],\\
		\quad (m,n) \neq (0,0)
	\end{split}
\end{align}
\nopagebreak%
\begin{vardesc}%
$W = (2d+1)²$
\end{vardesc}
The $i$th entry in the NGTDM is a sum of deviations from the mean of the center pixel, only concerning those pixels in the image which do not lie in the peripheral regions of width $d$.
\begin{align}
\label{eq:visualsi}
	s(i) = \begin{cases} 
		\displaystyle \sum_{p \in N_i}\left|i - \bar{A}_p\right|,&\; \text{there is a pixel with gray-tone } i\\
		0,&\;\text{otherwise}
	\end{cases}
\end{align}
\nopagebreak%
\begin{vardesc}%
$N_i$ -- the pixels with gray-tone $i$

$G_h$ -- the largest gray-tone
\end{vardesc}
The relative frequency, i.e. the probability of occurrence, of different gray-tones is calculated as:
\cite{Amadasun1989TexturalFeaturesCorrespondingToTexturalProperties}
\begin{align}
p_i &= \abs{N_i}/n, \notag\\
n &= (width-2d)(height-2d).\label{eq:visual_size}
\end{align}
Note that \eqref{eq:visual_size} allows a rectangular region of interest as opposed to the square regions used in the article by Amadasun and King, and that $n$ replaces $n^2$ in the formulas.

\subsubsection{Coarseness}
Coarseness is a measure of how rough a surface is, e.g. how large particles it is composed of.
\begin{align*}
f_{cos} = \left[\epsilon + \sum_{i=0}^{G_h} p_is(i)\right]^{-1}
\end{align*}
This is (inversely) a weighted sum of the deviations from the center pixels wrt the surrounding pixels. 
The small value $\epsilon$ is to cope with division by $0$.

\subsubsection{Contrast}
High contrast means the intensity difference between neighboring regions is large.
\begin{align*}
f_{con} &= \left[ \frac{1}{N_g(N_g-1)}\sum_{i=0}^{G_h}\sum_{j=0}^{G_h} p_i p_j (i-j)²\middle]\middle[\frac{1}{n}\sum_{i=0}^{G_h} s(i)\right]\\
\notag N_g &= \sum_{i=0}^{G_h}Q_i\\
\notag Q_i &= \begin{cases}
1 ,&\; \text{if } p_i \neq 0\\
0 ,&\; \text{otherwise}
\end{cases}
\end{align*}
\nopagebreak%
\begin{vardesc}%
$N_g$ -- the number of different gray-tones present in the image
\end{vardesc}
The first factor is used to reflect the dynamic range of gray scale weighted with the product of relative frequencies of the two gray-tone values under consideration.
The second factor increases with the amount of local variation in intensity.

\subsubsection{Busyness}
A busy texture is one where the spatial frequency of intensity changes are high.
\begin{equation*}
\begin{split}
f_{bus} = \left. \sum_{i=0}^{G_h} p_i s(i)  \middle/ \sum_{i=0}^{G_h} \sum_{j=i}^{G_h} ip_i - jp_j\right. \!\!,\\
  p_i\neq 0,\:p_j\neq 0
\end{split}
\end{equation*}
The numerator is a measure of the spatial rate of change in intensity, inversely related to coarseness. The denominator is a summation of the magnitude of differences between the different gray-tone values. 
This formula differs slightly from the one described in the article by Amadasun and King\cite{Amadasun1989TexturalFeaturesCorrespondingToTexturalProperties} --- I'm certain there's a typo in that formula making it always zero.

\subsubsection{Complexity}
Complexity means high information content. This could mean many primitives or patches, especially if they have different average intensity.
\begin{align*}
f_{com} = \sum_{i=0}^{G_h}\sum_{j=0}^{G_h} \left.\frac{\abs{i - j}}{n(p_i+p_j)}\Big( p_i s(i) + p_j s(j) \Big) \right.
\end{align*}
An elaborate description of this formula (and the others in this section) are found in the article by Amadasun and King\cite{Amadasun1989TexturalFeaturesCorrespondingToTexturalProperties}.

\subsubsection{Texture strength}
A strong texture is generally referred to as strong if its building blocks are easily definable and clearly visible. Such texture tend to look attractive. However a strong texture is difficult to define concisely\cite{Amadasun1989TexturalFeaturesCorrespondingToTexturalProperties}.
It is defined as
\begin{align*}
f_{str} = \frac{ \displaystyle \sum_{\substack{i=0}}^{G_h} \sum_{\substack{j=0}}^{G_h} \left( p_i + p_j \middle)\middle( i - j\right)^2}{ \displaystyle \epsilon + \sum_{i=0}^{G_h} s(i)}\!,
\qquad {p_i\neq 0,\:p_j\neq 0}.
\end{align*}
Where the numerator is a factor stressing the differences between intensity levels, and therefore may reflect intensity differences between adjacent primitives. The probabilities $p_.$ tend to be high for large primitives. The denominator would be small for coarse texture and high for busy or fine textures considering the definition in \eqref{eq:visualsi}.

\section{Fast 2D Convolution}
\label{sec:convolution}
Two-dimensional discrete convolution in the spatial domain is defined as
\begin{align*}
  (f * g)[n] \stackrel{\mathrm{def}}{=}\sum_{m=-\infty}^{\infty} f[m]\cdot g[n - m].
\end{align*}

By the Circular Convolution Theorem\cite{wikiCircularConvolution} this can instead be done in the frequency domain considering
\begin{align}
\mathcal{F}\{f*g\} = \mathcal{F}\{f\} \cdot \mathcal{F}\{g\}
\end{align}
\nopagebreak%
\begin{vardesc}%
$*$ -- the convolution operator

$\mathcal{F}\{\cdot\}$ -- the Fourier Transform (FT)
\end{vardesc}
First apply FT to image and to convolution kernel, then multiply the two matrices element-wise. To get the filtered image just apply inverse FT.

\begin{figure}
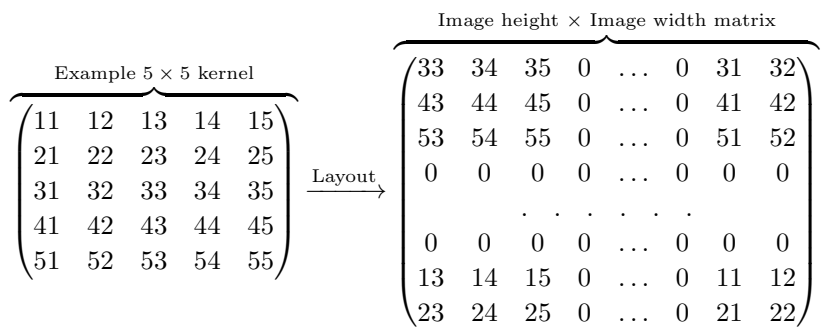

\begin{align*}
\overbrace{
\begin{pmatrix}
11 & 12 & 13 & 14 & 15\\
21 & 22 & 23 & 24 & 25\\
31 & 32 & 33 & 34 & 35\\
41 & 42 & 43 & 44 & 45\\
51 & 52 & 53 & 54 & 55\\
\end{pmatrix}
}^{\text{Example $5\times5$ kernel}}
\xrightarrow{\text{Layout}}
\overbrace{
\begin{pmatrix}
33 & 34 & 35 & 0 &\dots & 0 & 31 & 32 \\
43 & 44 & 45 & 0 &\dots & 0 & 41 & 42 \\
53 & 54 & 55 & 0 &\dots & 0 & 51 & 52 \\
0  & 0  &  0 & 0 &\dots & 0 & 0  &  0 \\
\hdotsfor[5]{8}\\
0  & 0  &  0 & 0 &\dots & 0 & 0  &  0 \\
13 & 14 & 15 & 0 &\dots& 0 & 11 & 12 \\
23 & 24 & 25 & 0 &\dots& 0 & 21 & 22
\end{pmatrix}
}^{\text{Image height $\times$ Image width matrix}}
\end{align*}
\caption{How to make sure the kernel wraps around the origin in frequency space}
\label{fig:CenterKernel}
\end{figure}

For this to work the kernel has to be placed in a matrix the same size as the image, wrapped around the origin\footnote{origin aka DC component, zero frequency}, which in FFTW is at position $\langle0,0\rangle$, like in figure~\ref{fig:CenterKernel}. 
Also, there are border cases in the image, it has to be padded with wraparound pixels.\cite{Convolution2DNVIDIA}

\pagebreak
\section{Scaling data}
\label{sec:scale}
Scaling is very important. If scaling is not applied to all features a feature with a larger numeric range may dominate others with smaller numeric range.
\begin{align*}
\displaystyle
range &= \max_i x_i - \min_i x_i\\
midrange &= \left(\displaystyle \max_i x_i + \min_i x_i\middle)\right/\!2\\
x'_i &= \begin{cases}
\displaystyle
\frac{x_i - midrange}{range/2},& range \neq 0\\
0,&range = 0
\end{cases}
\end{align*}\nopagebreak\begin{vardesc}%
$x_i$ -- feature value of example $i$

$i \in [0,\ell) \subset \Z^+$
\end{vardesc}

\chapter{Material and Methods}
\section{Material}
Blood samples were taken from four individuals. The cells were photographed on a \cellavision~DM-96. 
The width of the images lies in the range $[119,267]$. The height of the images lies in the range $[119,258]$. On average an image is about $139\times139$ pixels. This correspond to about $13.7$\ \hbox{$\umu \mathrm{m}$}.

The cells are normal, e.g. there are no cancer cells or malaria infected cells.
 There are very few (2) blast cells indicating the only possible cancer type would be lymphoma, i.e. a cancer in the lymph nodes.

The cells were classified on the \cellavision\ DM-96 and its result was taken as {\em ground truth}. 
The machine is 90\% to 95\% correct depending on the individual.
The cell types of the data set are given in table~\ref{tab:celltypes}. 
Typical relative frequencies of the cells are found in table~\ref{tab:cellabundance}. 
Typical images of some common cells are found in figure~\ref{fig:celltypes}.
\begin{table}
  \begin{center}
    \begin{tabular}{rl}
  		\toprule%
      {\scshape Class No.} & {\scshape Class Name} \\
      \midrule%
			1 & neutrophil granulocytes, segmented \\
			6 & neutrophil granulocytes, band  \\
			2 & eosinophil granulocytes\\
			3 & basophil granulocytes\\
			4 & lymphocytes \\
			7 & lymphocytes, variants \\
			5 & monocytes \\
			9 & myelocytes \\
			10 & meta-myelocytes \\
			11 & blast, immature cell \\
			21 & artifacts \\
			24 & broken cell \\
			25 & thrombocytes (platelets)\\
			29 & clots of thrombocytes \\
			\bottomrule%
		\end{tabular}
		\caption{Cell types classified in the data set}
	\end{center}
\label{tab:celltypes}
\end{table}

From the set of images of the cells a range of descriptors, or {\em features}, were extracted. 
A set of features extracted from a single image, called {\em instance} or {\em example}, is denoted $\x$ and the space of all possible features is denoted $\X$.

A Support Vector Machine (SVM) was trained using the set of features described.

\section{Implementation details}
\subsection{Support Vector Machine}
The SVM was written in C++ within the {\em Boost C++ Libraries} framework.
The Gram matrix $\vect G$, defined in \eqref{eq:GramMatrix}, the output of the kernel function, is cached in memory to dramatically reduce running time.

\subsubsection{A Stochastic Gradient Ascent Variant}
Stochastic gradient ascent differs from ordinary gradient ascent in that the coefficients $\alpha_i$ updated are used right away, instead of in the next iteration.
In this project a variant of the stochastic gradient ascent method of training a SVM were implemented.

The coefficients $\vect \alpha_{KKT}$ that invalidate the Karush-Kuhn-Tucker\ (KKT) conditions are selected first for update.
They are likely the ones that will affect the solution most rapid.
When these satisfies the KKT conditions, or when no progress has been made in some iterations, the greater problem of updating all coefficients $\vect \alpha$ is considered.

\subsubsection{Multiclass SVM}
I use the one-against-the-rest method\cite{HsuLinMultiClass} because it is the simplest and it has similar precision to the latter two\cite{Vapnik,HsuLinMultiClass}.
The latter two are however faster to train because they can train all the classifiers at once.\cite{Nello}

\subsection{Features}
\subsubsection{Scalable Color Descriptor}
In MPEG-7 the 3D color histogram bins are reduced in size by truncation and encoding (see~\ref{sec:scalablecolor}). 
To release the SVM from this hassle it receives the values as ordinary real values representing the relative frequency of color channel values.
The bounded time complexity to calculate this descriptor is $O(3W\!H)$.

\subsubsection{Color Structure Descriptor}
This is implemented by calculating a histogram for each structuring element and then summing over all structuring elements
\begin{align}
h(m)=\sum_{i=1}^{\frac{W-8K}{K}}\sum_{j=1}^{\frac{H-8K}{K}}\min\{1,h_{s_{i,j}}(c_m)\}
\end{align}
\nopagebreak%
\begin{vardesc}$m$ -- bin index in the final histogram

$c_m$ -- quantized color level

$h_{s_{i,j}}$ -- histogram for structuring element $\langle i,j\rangle$
\end{vardesc}
Calculating this descriptor is much more expensive than Scalable Color Descriptor described in section~\ref{sec:scalablecolor}, 
$O(\frac{(w-8k)(h-8k)}{k} 8^2)$ for each channel, this is more than a 30-fold increase on a $640\times480$ image compared to the above.
\subsubsection{Color Layout Descriptor}
The Discrete Cosine Transform of type DCT-II is calculated using the software library FFTW3 (Fastest Fourier Transform in the West).
The zigzag scanning order described in figure~\ref{fig:zigzag} is implemented as an C++ STL iterator using the simple algorithm presented in listing~\ref{lst:zigzag}.
A wider low pass band is used than in MPEG-7. 
The 10 first Y (6 in MPEG-7), the 5 first of U and V (3) coefficients are extracted.
\lstset{language=C}
\lstset{tabsize=2}
\lstset{frameround=fttt}
\begin{lstlisting}[float,frame=trBL,caption=Simplified source for the implemented zigzag order on a length$\times$length square matrix,label=lst:zigzag,captionpos=b]
  x = 0; y = 0; forward = true;
  value_type get_current() { return source(x,y); }
  void next() {
	  if (forward)
			if (y < length-1) {
				y ++; x --;
				if (x < 0) { 
					x = 0;
					forward = false;
				} 
			} else
				if (y == length-1) {
					x ++;
					forward = false;
				}
		else
			if (x < length-1) {
				x ++; y --;
				if (y < 0) {
					y = 0;
					forward = true;
				}
			} else
				if (x == length-1) {
					y ++;
					forward = true;
				}
  }
\end{lstlisting}
\subsubsection{Homogeneous Texture Descriptor}
By symmetry the filter might as well be rotated instead of the image and since that is more efficient that is what is done.
The bandwidth $b$ is set to 1 octave by relation \eqref{eq:bandwidth} and setting $\sigma = \sigma_x$
\begin{align}
\label{eq:bandwidth}
\frac{\sigma}{\lambda} &= \frac{1}{\pi}\sqrt{\frac{\ln2}{2}} \frac{2^b + 1}{2^b - 1} \approx 0.5622
\end{align}
In MPEG-7 rotation invariance in this descriptor is achieved by rotating the features in the direction of the dominant direction. 
This is not implemented in this project.

\begin{figure}
	\begin{center}
		\subbottom[$\theta = 0°$]{ \includegraphics{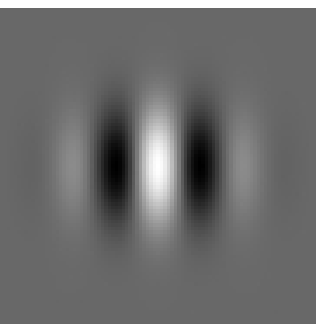} }
		\subbottom[$\theta = 36°$]{ \includegraphics{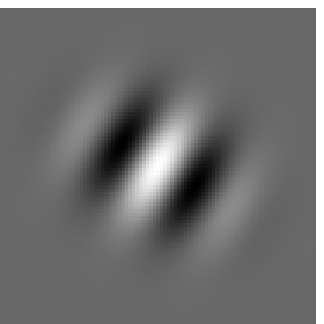} }
		\subbottom[$\theta = 72°$]{ \includegraphics{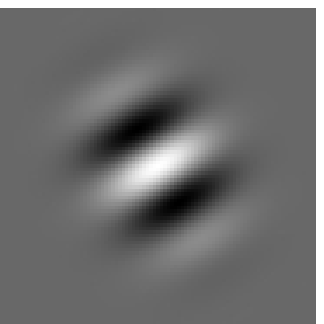} }
		\\
		\subbottom[$\theta = 108°$]{ \includegraphics{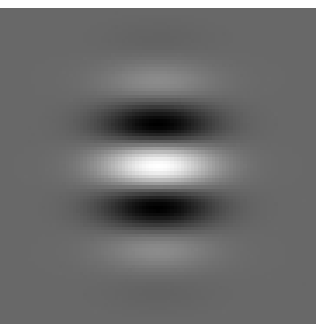} }
		\subbottom[$\theta = 144°$]{ \includegraphics{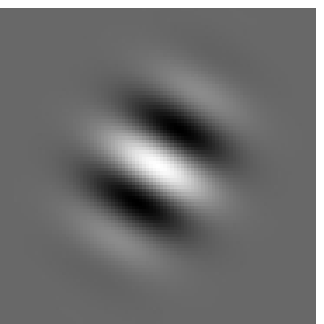} }
		\subbottom[$\theta = 180°$]{ \includegraphics{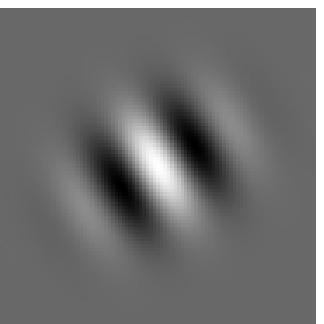} }
		\caption{Gabor Filter bank at scale = $S-1$ at different orientations. Gray areas are the ones with zero magnitude, darker is negative, lighter is positive}
		\label{fig:gaborkernels}
	\end{center}
\end{figure}

In figure~\ref{fig:gaborkernels} images of the Gabor wavelet filter bank kernels of different orientations are presented.

\subsubsection{Neighborhood Gray-Tone Difference Matrix}
The $\bar{A}$ used in the Neighborhood Gray-Tone Difference Matrix \eqref{eq:NGTDMAbar} can be divided into subproblems which do not need to be calculated every time.
By keeping the center value $(m,n)=(0,0)$ in the sum (not writing out normalization)
\begin{align*}
	A'(k,l) &= \sum_{m=-d}^d\sum_{n=-d}^d f(k+m,l+n),\\*
	\intertext{it can also be written as}
	A'(k,l) &= \begin{cases}
		\begin{split}
			\underbrace{A'(k, l-1)}_{\text{above}} +\phantom{\qquad\text{or as }}\\*
			\sum_{m=-d}^d f(k+m,l+d) - f(k+m,l-d-1) \qquad \text{or as }
		\end{split}
		\\
		\begin{split}
			\underbrace{A'(k-1, l)}_{\text{to the left}} +\\*
			\sum_{n=-d}^d f(k+d,l+n) - f(k-d-1,l+n).
		\end{split}
	\end{cases}
\end{align*}
Given the value above or the value to the left the others can be calculated faster.

To find all $\bar{A}$ first fill in a table with all $A'$, from left to right, top-down. 
Then for all positions remove the center value and make sure the accumulated value is correctly normalized.
The time complexity is thereby reduced from $O(d^2)$ per pixel to $O(d)$ per pixel.

\subsection{Convolution}
\label{sec:convolution2}
Using the method for convolution described in section~\ref{sec:convolution} is much more efficient than the naive approach of doing the calculations in the spatial domain.
It reduces the complexity from $O(K²)$ per pixel, where $K$ is the size of the convolution kernel, to $O(\log N)$, where the image is $N\times N$ in size and $N = 2^k, k\in \Z^+$. 
The last requirement make sure that the much more efficient Fast Fourier Transform (FFT) can be used instead of a normal Discrete Fourier Transform (DFT).

With the largest kernel used, $K² = 91² = 8281$, and a $1000\times 1000$ image, $\log 1000 \approx 6.9$, a thousandfold speed-up can be achieved.

These figures are however for FFT on matrices of size $N = 2^k$. 
Padding to the next larger 2-power is not implemented since the software library used for FFT, called FFTW\footnote{Heavily used library with an impressing architecture, used in e.g. Matlab} (Fastest Fourier Transform in the West) supports other sizes too and still provides great speed.

\subsection{Data View}
\begin{figure}
	\begin{center}
		\includegraphics[scale=0.75]{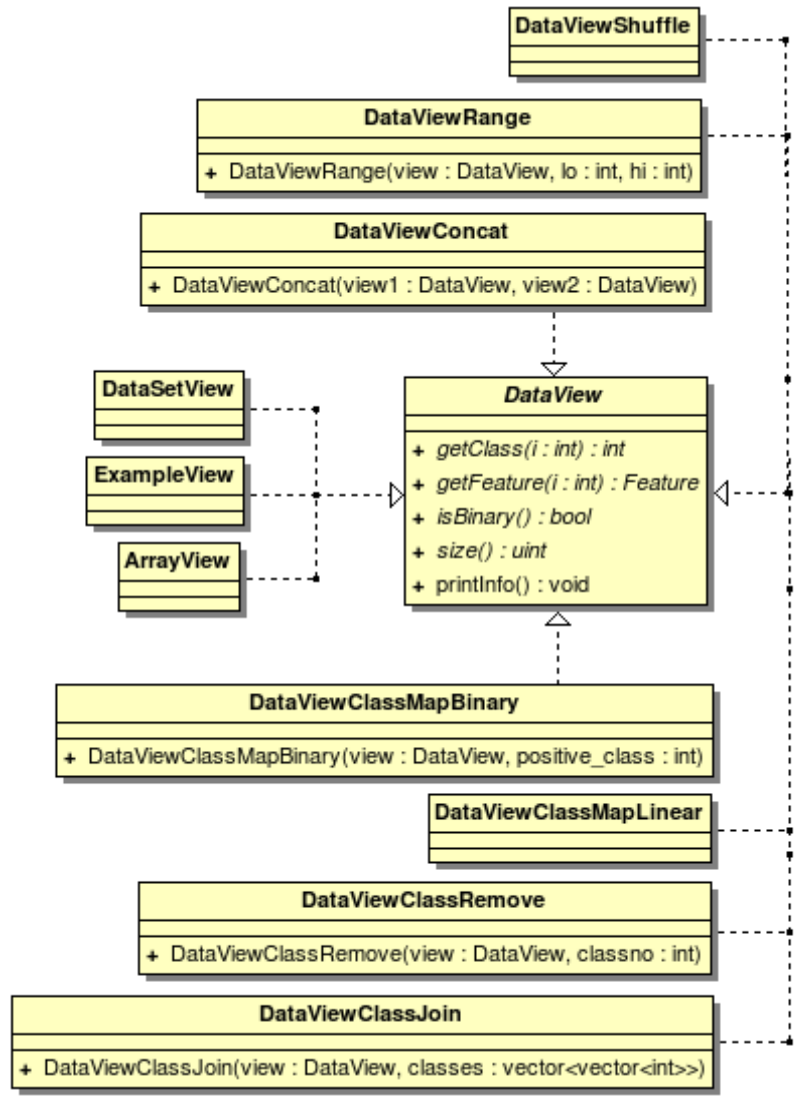}
		\caption{Abstract class (interface) to data views and their realizations}
		\label{fig:DataView}
	\end{center}
\end{figure}

The classifiers view data. 
Rather than giving them the data structure holding data directly an abstraction was built named {\ttfamily DataView}. 
The abstraction was realized in 11 classes which are found together with their base abstract class in figure~\ref{fig:DataView}.
The derived classes can all be used transparently releasing the classifier and data set loader from the tasks of the views.

\pagebreak

These three views below contain pointers to the real data. 

\vspace{1em}
\hspace{3.05cm}%
\begin{minipage}[t]{0.705\textwidth}
\begin{itemize}
  \item[\ttfamily DataSetView] view of data represented by a {\ttfamily DataSet} instance
  \item[\ttfamily ExampleView] view of data represented by a vector of {\ttfamily Example} instances
  \item[\ttfamily ArrayView] view of data from an {\ttfamily boost::Array}, convenient for the unit tests concerning views
\end{itemize}
\end{minipage}
\vspace{1.5em}\\

The views below contain other views and just map their values. 
They are often chained together to get the wanted view.

\vspace{1em}
\hspace{3.05cm}%
\begin{minipage}[t]{0.705\textwidth}
\begin{itemize}
  \item[\ttfamily DataViewScaled] view the features as if they were in the range $[-1,1]$, avoids feature-wise bias, see section~\ref{sec:scale}
  \item[\ttfamily DataViewRange] selected only a subset of the examples, used in e.g. cross-validation
  \item[\ttfamily DataViewConcat] view two views as if they were one, also used in cross-validation
  \item[\ttfamily DataViewShuffle] shuffle the order of examples. It is of course not wanted to split an ordered set and train on the first part and test on the other, a class may then be present only in the latter
  \item[\ttfamily DataViewClassMapLinear] if e.g. only classes $\{0,3,42,\ldots\}$ exists it is convenient if they can be represented by $\{0,1,\ldots\}$
  \item[\ttfamily DataViewClassMapBinary] one class given is said to be positive, all other is said to be negative. Used in multiclass classifier
  \item[\ttfamily DataViewClassJoin] join groups of classes into new classes
  \item[\ttfamily DataViewClassRemove] view with a class removed
\end{itemize}
\end{minipage}

\chapter{Experimental Setup and Results}
\section{Experimental Setup}
The \cellavision~DM-96 machine achieves an error rate of approx.~5-10\% depending on individual. 
Thus there are errors in the ground truth.

I have divided the problem in two parts. 
\begin{itemize}
	\item {\em the primary problem} --- the SVM should classify all classes present in the data set.
	\item {\em the simplified problem} --- some classes are merged and others are removed.
\end{itemize}

\subsection{Performance test method}
In both cases 2-fold cross-validation is used to test performance. 
This means that two models will be trained.
In the first, half the data set is the training set and the other half is the test set. 
In the other, the roles of the subsets are swapped. 
This way both halves will act as both training and test sets.

\subsection{Description of the simplified problem}
Class 1 and 6, Neutrophil granulocytes, segmented and band variants are merged to form class 30.
Even human experts have approx. 25\% error rate on these. 
It is often more a matter of opinion than of objective decision.

Class 4 and 7, Lymphocytes and their variants, are joined. 
The variants are rather uncommon, there are only 8 instances in the dataset, compared to 160 of Lymphocytes. 
Due to the skew distribution these are merged to form class 31.

The following classes are removed.
Class 0 are unidentified objects, it is a very heterogeneous group but there are only 6 of them.
Class 21 are artifacts, random garbage, they are removed.
Class 24 are broken cells, there are only 7 of them.
Class 25 and 29 are thrombocytes and clots of them, i.e. platelets. 
Since they aren't even white blood cells they are removed.
Class 11, called {\em blast} is a kind of immature cell which would be interesting to classify but there are only two of them so they are removed as well.
Class 9 and 10 are myelocytes and meta-myelocytes, which are a development stage of different granulocytes. 
There can be e.g. eosinophilic myelocytes and basophilic myelocytes. 
In the dataset they are also too rare to train a general classifier. 
There are only a total of 4 myelocytes in this heterogeneous group.
All classes that are left are presented again in table~\ref{tab:celltypessimpl}.
\begin{table}
	\begin{center}
		\begin{tabular}{rl}
			\toprule%
			{\scshape Class No.} & {\scshape Class Name} \\
			\midrule%
			30 (1+6) & neutrophil granulocytes \\
			2 & eosinophil granulocytes\\
			3 & basophil granulocytes\\
			31 (4+7) & lymphocytes and variants\\
			5 & monocytes \\
			\bottomrule%
		\end{tabular}
		\caption{Cell types left in the simplified problem}
		\label{tab:celltypessimpl}
	\end{center}
\end{table}

\section{Results}
\subsection{Primary Problem}
The error rate in the primary problem is 10.8\%. 
The type of kernel function that was the most successful was the Polynomial kernel. 
This is compared to the slightly better result using libSVM, 9.6\%.
See table~\ref{tab:primresult}.

Most confusion occurs between classes 1 (segmented neutrophil granulocytes) and 6 (band neutrophil granulocytes).
Much confusion is also present when recognizing class 3 (basophil granulocytes) --- they are often (2 of their total of 7) misclassified as class 1 (segmented neutrophil granulocytes), which is a very large group.

\subsection{Simplified Problem}
In the simplified problem the error rate is 3.1\%.
Also in this problem the most successful kernel was the Polynomial kernel. 
This is compared to the better result using libSVM, 2.3\%.
See table~\ref{tab:simplresults}.

In the simplified problem most confusion (by number) occurs between class 5 (monocytes) and the new class 31 (lymphocytes).
By percentage the largest confusion occurs between class 30 (segmented and band neutrophil granulocytes) and class 3 (basophil granulocytes).
Class 3 have only 8 instances of which 3 were misclassified as 30.

\begin{table}
	\begin{center}
		\newcolumntype{K}{R[,][.]{5}{1}}
		\begin{tabular}{rKKKl}
			\toprule%
			\multicolumn{5}{c}{\scshape Implemented SVM Results} \\
			\multicolumn{1}{r}{\scshape Kernel Type} & \multicolumn{3}{c}{\scshape Error Rate (\%)} & \multicolumn{1}{l}{\scshape Parameters}\\
			\midrule%
			&\multicolumn{1}{c}{\scshape Total} & \multicolumn{1}{c}{\scshape Max} & \multicolumn{1}{c}{\scshape Min} &  \\
			\midrule%
			RBF with $L^2$ norm  & 11,5385 & 12,0192 & 11,0577  & $\sigma²=20$\\
			RBF with $L^2$ norm  & 11,5385 & 12,0192 & 11,0577  & $\sigma²=22$\\

			Polynomial & 11,7788 & 12,5    & 11,0577  & $d = 2$ \\
			Polynomial & 11,0577 & 11,5385 & 10,5769  & $d = 3$ \\
			Polynomial & 11,5385 & 12,0192 & 11,0577  & $d = 4$ \\
			Polynomial & 10,8173 & 11,0577 & 10,5769 & $d = 5$\\
			Polynomial & 11,2981 & 12,0192 & 10,5769 & $d = 6$ \\
			\midrule
			\multicolumn{5}{c}{\scshape libSVM Results} \\
			\midrule
			RBF & 9,5923 &\skiptwo & $C=512, \gamma^{-1} = 8192$\\
			\bottomrule%
		\end{tabular}
		\caption{SVM cell classifier results for the primary problem}
		\label{tab:primresult}
	\end{center}
\end{table}
\begin{table}
	\begin{center}
		\begin{tabular}{rrccccccccccc}
			\toprule
			\multicolumn{13}{c}{\scshape Number of Confusions}\\
			\midrule
			& \multicolumn{11}{c}{\scshape Guessed Class}\\
			{\scshape Class} & $(n)$ &          0&         1&         2&         3&         4&         5&         6&         7&        11&        21&        24\\
			\midrule
			0& (4)&   $\cdot$ &   $\cdot$ &   $\cdot$ &   $\cdot$ &   $\cdot$ &   $\cdot$ &   $\cdot$ &   $\cdot$ &   $\cdot$ &   $\cdot$ &   $\cdot$\\
			1& (205)&   $\cdot$ &   $\cdot$ &   $\cdot$ &   $\cdot$ &   $\cdot$ &   $\cdot$ &         1 &   $\cdot$ &   $\cdot$ &   $\cdot$ &   $\cdot$\\
			2& (14)&         1 &   $\cdot$ &   $\cdot$ &   $\cdot$ &   $\cdot$ &   $\cdot$ &   $\cdot$ &   $\cdot$ &   $\cdot$ &   $\cdot$ &   $\cdot$\\
			3& (7)&   $\cdot$ &         2 &   $\cdot$ &   $\cdot$ &   $\cdot$ &   $\cdot$ &   $\cdot$ &   $\cdot$ &   $\cdot$ &   $\cdot$ &   $\cdot$\\
			4& (104)&   $\cdot$ &   $\cdot$ &   $\cdot$ &   $\cdot$ &   $\cdot$ &   $\cdot$ &   $\cdot$ &         2 &   $\cdot$ &         1 &   $\cdot$\\
			5& (32)&   $\cdot$ &   $\cdot$ &   $\cdot$ &   $\cdot$ &         2 &   $\cdot$ &   $\cdot$ &         1 &   $\cdot$ &   $\cdot$ &   $\cdot$\\
			6& (12)&         1 &         6 &   $\cdot$ &   $\cdot$ &   $\cdot$ &         1 &   $\cdot$ &   $\cdot$ &   $\cdot$ &   $\cdot$ &   $\cdot$\\
			7& (6)&   $\cdot$ &   $\cdot$ &   $\cdot$ &   $\cdot$ &         1 &         1 &   $\cdot$ &   $\cdot$ &   $\cdot$ &   $\cdot$ &   $\cdot$\\
			11& (1)&   $\cdot$ &   $\cdot$ &   $\cdot$ &   $\cdot$ &         1 &   $\cdot$ &   $\cdot$ &   $\cdot$ &   $\cdot$ &   $\cdot$ &   $\cdot$\\
			21& (31)&   $\cdot$ &   $\cdot$ &         1 &   $\cdot$ &   $\cdot$ &   $\cdot$ &   $\cdot$ &   $\cdot$ &   $\cdot$ &   $\cdot$ &   $\cdot$\\
			24& (1)&   $\cdot$ &   $\cdot$ &   $\cdot$ &   $\cdot$ &   $\cdot$ &   $\cdot$ &   $\cdot$ &   $\cdot$ &   $\cdot$ &         1 &   $\cdot$\\
			\bottomrule
		\end{tabular}
		\caption{Confusion Matrix for the primary problem}
		\label{fig:result_confusion}
	\end{center}
\end{table}

\begin{table}
	\begin{center}
		\newcolumntype{K}{R[,][.]{5}{1}}
		\begin{tabular}{rKKKl}
			\toprule
			\multicolumn{5}{c}{\scshape Implemented SVM Results} \\
			\multicolumn{1}{r}{\scshape Kernel Type} & \multicolumn{3}{c}{\scshape Error Rate (\%)} & \multicolumn{1}{l}{\scshape Parameters}\\
			\midrule
			&\multicolumn{1}{r}{\scshape Total} & \multicolumn{1}{r}{\scshape Max} & \multicolumn{1}{r}{\scshape Min} \\
			\midrule
			RBF        & 3,64109 & 4,42708 & 2,86458 & $C=128, \sigma^2=16$ \\
			RBF        & 3,25098 & 4,16667 & 2,34375 & $C=512, \sigma^2=128$ \\
			Polynomial & 3,12094 & 3,38542 & 2,864 & $d=3$ \\
			Polynomial & 3,51105 & 3,90625 & 3,125 & $d=5$ \\
			\midrule
			\multicolumn{5}{c}{\scshape libSVM results} \\
			\midrule
			RBF        & 2,470  & \skiptwo & $C=8, \gamma^{-1}=128$\\
			Polynomial & 2,3407 & \skiptwo & $C=8, \gamma^{-1}=128, d=3$ \\
			Polynomial & 3,5111 & \skiptwo & $C=8, \gamma^{-1}=128, d=5$ \\
			\bottomrule
		\end{tabular}
		\caption{SVM cell classifier results for the simplified problem}
		\label{tab:simplresults}
	\end{center}
\end{table}
\begin{table}
	\begin{center}
		\begin{tabular}{rrccccc}
			\toprule
			\multicolumn{7}{c}{\scshape Number of Confusions}\\
			\midrule
			& \multicolumn{5}{c}{{\scshape Guessed class}}\\
			{\scshape Class} & $(n)$ &          2&         3&         5&        30&        31\\
			\midrule
			2& (20)&   $\cdot$ &   $\cdot$ &   $\cdot$ &   $\cdot$ &   $\cdot$\\
			3& (8)&   $\cdot$ &   $\cdot$ &   $\cdot$ &         3 &   $\cdot$\\
			5& (56)&   $\cdot$ &   $\cdot$ &   $\cdot$ &   $\cdot$ &         2\\
			30& (517)&   $\cdot$ &   $\cdot$ &         1 &   $\cdot$ &   $\cdot$\\
			31& (168)&   $\cdot$ &   $\cdot$ &         7 &   $\cdot$ &   $\cdot$\\
			\bottomrule
		\end{tabular}
		\caption{Confusion matrix for the simplified problem}
	\end{center}
\end{table}

\chapter{Discussion}
The accuracy achieved in the primary problem was 89.2\% and in the simplified problem 96.9\%. 
I regard these results as good when compared to \cellavision~DM-96's result of the primary problem, 90-95\%. 
One has to consider that there are errors in the ground truth misleading the SVM.
Thus, it is uncertain whether the results are better than the DM-96 or worse. 
Because the DM-96 has an error rate of about 5-10\% a 0\% error rate in the primary problem would mean something like 5-10\% error, while a 5\% error could possibly mean 0-15\% error.

I conclude that using the combination of MPEG-7 descriptors and visual texture features in combination with SVM to classify cells is good but need further investigation to find out {\em how} good.
A more comprehensive study could investigate whether a set of SVM or ANN variants perform better on the set of features implemented or on the set of features developed at \cellavision.

I would like to stress that using a SVM instead of an Artificial Neural Network as in the \cellavision~DM-96 machine is more statistically rigor --- Confidence intervals of the classifier can be found, which to my knowledge is impossible in ANNs.
In medicine it is important to know the strength of the method used.

It would be very interesting to test the features on the real training set they have developed at \cellavision.
The company has a training set of thousands of cells classified by field experts. 
Some cell images required five experts to be certain of the cell type.
Without the errors in the ground truth the results could possibly compete with the \cellavision machine.

The result of the primary problem states that the most confused instances are those that are guessed to be a segmented neutrophil (class 1) but that are a band neutrophil (class 6) in the ground truth. 
These two often look very similar, humans often have different opinions about which class cells are.
Also the \cellavision~DM-96 have problems with these classes indicating that there are several errors in the ground truth.
The errors in ground truth probably mislead the SVM. 
There are only 12 cell images of class 6 of which some have the wrong class and there are 205 images of class 1, of which not all are truly class 1.
This situation pushes bias to the larger class.

In the result of the simplified problem most confusion occur between the monocytes (class 5) and the lymphocytes (class 31).
This was expected as they are very hard to classify for both humans and the \cellavision~DM-96.
Late in writing this thesis I discovered that there is great discrepancy in size of these two types of cells.
The discrepancy indicates that the size of the cells could be used as a feature too.

\subsection*{Practical Use}
Even the simplified problem would give useful information when applied to medicine.
Standard measures used in diagnosis involve counting the total number of white blood cells, leukocytes, determining the distribution of lymphocytes and granulocytes and determining the number of monocytes.

Malaria infected and cancer cells look different compared to healthy blood cells. 
It would be interesting to test the features on these kind of cells to be able to classify them as well.

\subsection*{Runtime Performance}
To increase cache performance in the Color Structure Descriptor (section~\ref{sec:ColorStructureDescriptor}) it would be wise to first extract all sub-samples i.e. the representative color for each $K\times K$ area as the other pixels aren't used.
They will otherwise quickly fill up the cache during memory pre-fetch. 
Now, the sub-samples are viewed using a sub sampling view present in Generic Image Library (boost::gil), contributed to Boost by Adobe.
The views in GIL are virtual, meaning they only keep information about offset calculations --- no data is duplicated.

The 2D convolution was first done in the spatial domain but I soon realized it was way to slow with my bigger Gabor filter kernels of which the largest are $91²$ pixels big. Instead the calculations are done in the frequency domain which is much faster, see sections~\ref{sec:convolution}~and~\ref{sec:convolution2}.

To improve performance of the Gabor Wavelet Filter further the kernels should of course be kept in memory when generating features of many images, however they are not.

To improve SVM training performance the Gradient Ascent training algorithm must be replaced or at least improved. 
The algorithm implemented divide the problem into a subproblem where the coefficients violating the KKT conditions are first optimized. 
This is a heuristic called {\em chunking} in the literature\cite{Nello}. 
By using this, fewer elements of the Gram matrix, and their corresponding support vectors, need to be kept in memory. 
This is something I don't take advantage of because I had enough memory for my purposes.
By refining chunking into {\em decomposition} where a fixed size chunk is optimized, more data points can be used and convergence speed is increased.
The Sequential Minimization Optimization (SMO) takes decomposition to the extreme and optimizes only two coefficients at a time and can thereby make sure that the KKT condition, $\sum_{i=1}^\ell\alpha_iy_i=0$, is always true. LibSVM uses a variant of this approach and it offer great performance.\cite{CC01aLIBSVM,LIBSVMDimensionalityReductionViaSparseSupportVectorMachines}

\subsection*{Beyond Gabor Filters}
If modeling human brains is the objective, considering other approaches than the Gabor wavelet would be interesting.
A type of neurons in the first visual cortex, called {\em simple cells}, have been recorded from monkey and cat.
The recordings and the elaborate analytical discussion in an article by Wallis show that both {\em difference of Gaussian$\times$Gaussian} (DoGG) and Cauchy functions model cortical cells better than Gabor wavelets for the measured parameters.\cite{Wallis_Linear_Models_Of_Simple_Cells_Mammal_Vision_Model}
In an article by Ashour et al. three other types of transforms are suggested --- ridgelets, curvelets and contourlets.\cite{Ashour2008SupervisedTextureClassificationUsingSeveralFeaturesExtractionTechniquesBasedOnAnnAndSvm}
Perhaps they can show increased performance.

\renewcommand\bibname{References}
\bibliography{article}


\appendix
\chapter{Software Usage}
The software produced in this project can be found at
\begin{itemize}
	\item \verb|http://tobbe.nu/pub/2008/cell.morph.mpeg7.svm/|
\end{itemize}
The software has only been tested on an Ubuntu Linux system.
However, the software is written in portable C99 C++ and should work on all *nix platforms that can supply the dependencies, perhaps even under {\em cygwin} under MS Windows.
The dependencies are
\begin{itemize}
	\item C99 compliant C++ compiler (GNU g++ tested)
	\item Boost C++ Libraries, \texttt{http://www.boost.org/}
	\item FFTW3 (Fastest Fourier Transform in the West 3), \texttt{http://www.fftw.org/}
	\item GSL (GNU Scientific Library), \texttt{http://www.gnu.org/software/gsl/}
	\item libjpeg
	\item libpng
\end{itemize}

Below is a brief overview on how to use the most important programs in the software package.
There are other programs in the package but they are mostly related to testing.
\pagebreak
\section{\texttt{train} -- Train a model}
This is the program where most processing is done. It can 
\begin{itemize}
 \item train a model from a dataset
 \item test a model with a dataset
 \item load and/or save a model from/to a file
 \item perform cross-validation
\end{itemize}
Here is the syntax of the program \texttt{train}
\begin{verbatim}
        MAIN ::= (MAIN_HELP | MAIN_DO)
   MAIN_HELP ::= ./train [-h]
     MAIN_DO ::= ./train MODE DATASET 
                         MODEL_PARAMS SAVE_MODEL
        MODE ::= LOAD_MODEL XVALIDATION
  LOAD_MODEL ::= -l MODEL.model
 XVALIDATION ::= -f N_FOLDS
     N_FOLDS ::= 1 | INTEGER
     DATASET ::= -d INTEGER
MODEL_PARAMS ::= -k KERN -p KERN_PARAM 
                 -C DOUBLE -g GAP_TOL -m TERM
        KERN ::= KERN_LIST | KERN_TYPE
   KERN_LIST ::= 0
   KERN_TYPE ::= 1 | 2 | 3 | 4 | 5 | 6 | 7
  KERN_PARAM ::= DOUBLE
     GAP_TOL ::= DOUBLE
        TERM ::= BITMASK
     BITMASK ::= 1 | 2 | 3
  SAVE_MODEL ::= -o MODEL.model
\end{verbatim}
Both cross-validation and saving of a model can be performed at the same time if wanted.
However, this will mean that \texttt{train} will create one model for each fold but it is just the last one that will be saved. 
If cross-validation is not wanted pass one (\texttt{-f 1}) fold.
The double precision floating point number passed with \texttt{-C} is a number used in the classifier, it is related to the KKT conditions.
The gap tolerance is also a double precision floating point number which is used as a convergence criterion. 
It is the allowed gap between the primal and dual objective function, the feasibility gap, which should be a small number. 
The default gap is set to $10^{-3}$. 
The \texttt{m} terminator is a bit-mask which control when a classifier is considered optimal, i.e. when training will stop.
The feasibility gap constraint is not used if \texttt{-m 2} is passed, i.e. when the first bit (1) is zero. 
The primary training terminator bit is 2 which means that all KKT conditions must be satisfied to terminate training. 
The default of 3 means that both these conditions must be satisfied.

\section{\texttt{cellfeatures} -- Generate examples from the cell database}
To generate features from all pairs of (image,ground truth class) in the cell database the program \texttt{cellfeatures} is used.
The file \texttt{cellfeatures.data} is backed up before writing the features generated to it.
This file can be used by the program \texttt{train}.
\begin{verbatim}
       MAIN ::= (MAIN_HELP | MAIN_DO)
  MAIN_HELP ::= ./cellfeatures
    MAIN_DO ::= ./cellfeatures DB
\end{verbatim}

\section{\texttt{jpeg\_genfeature} -- Feature generation from images}
To generate a set of features from image(s) the program called \texttt{jpeg\_genfeature} is used.
It generate a set of features that can be classified later with \texttt{predict}. 
\begin{verbatim}
     MAIN ::= (MAIN_HELP | MAIN_DO)
MAIN_HELP ::= ./jpeg_genfeature -?
  MAIN_DO ::= ./jpeg_genfeature CROPIMAGE* -o FEATURESET.feat
CROPIMAGE ::= -i IMAGE.jpeg [-x left -y top -w width -h height]
\end{verbatim}

\section{\texttt{predict} -- Predicting a set of features}
To predict a set of features, generated by \texttt{jpeg\_genfeature}, the program called \texttt{predict} is used.
It needs a previously trained model generated by \texttt{train}. 
\begin{verbatim}
     MAIN ::= (MAIN_HELP | MAIN_DO)
  MAIN_DO ::= ./predict -l MODEL.model -f FEATURESET.feat
MAIN_HELP ::= ./predict -?
\end{verbatim}

\section{\texttt{extractcelltype} -- Extract a class of images from the cell database}
To extract a specific class (as classified by \cellavision~DM-96) from the cell database, the program called \texttt{extractcelltype} is used.
\begin{verbatim}
       MAIN ::= (MAIN_HELP | MAIN_DO)
  MAIN_HELP ::= ./extractcelltype
    MAIN_DO ::= ./extractcelltype CLASS DB
      CLASS ::= INTEGER
        DB  ::= ALLXMLFILES | (XMLFILE ' ')*
ALLXMLFILES ::= '.'
\end{verbatim}
\section{\texttt{extractcellid} -- Extract given instances from the cell database}
To extract given instances from a list of id numbers, the program called \texttt{extractcellid} is used.
\begin{verbatim}
       MAIN ::= (MAIN_HELP | MAIN_DO)
  MAIN_HELP ::= ./extractcellid
    MAIN_DO ::= ./extractcellid IDLIST DB
     IDLIST ::= (INTEGER ' ')* 'x'
\end{verbatim}

\section{\texttt{extractcellinfo} -- Extract statistics of instances from the cell database}
To extract statistics about size, resolution and number of instances of a specific class or of all classes the program called \texttt{extractcellinfo} is used.
\begin{verbatim}
       MAIN ::= (MAIN_HELP | MAIN_DO)
  MAIN_HELP ::= ./extractcellinfo
    MAIN_DO ::= ./extractcellinfo CLASS DB
      CLASS ::= CLASS_ALL | CLASS
  CLASS_ALL ::= '-1'
\end{verbatim}

\section{\texttt{tolibsvm} -- Save cell features in libSVM format}
This program load the features saved in \texttt{cellfeatures.data} and dump them in libSVM format on standard output. 
It takes no parameters.
\begin{verbatim}
./tolibsvm > cellfeatures.data.libsvm
\end{verbatim}

\end{document}